\documentclass[twocolumn,showpacs,preprintnumbers,amsmath,amssymb]{revtex4}

\usepackage{graphicx}
\usepackage{dcolumn}
\usepackage{bm}

\begin{document}


\title{Collective Modes and Stability of Bose-Fermi Mixtures with a BCS-BEC Crossover}

\author{Hiroyuki Shibata}
\email{shibata@kh.phys.waseda.ac.jp}
\author{Nobuhiko Yokoshi}
\author{Susumu Kurihara}
\affiliation{Department of Physics, Waseda University, Okubo, Shinjuku, Tokyo 169-8555, Japan}

\date{\today}

\begin{abstract}
	We investigate an ultracold Bose-Fermi mixture with a Feshbach resonance between two hyperfine states of fermions. Using a functional integral method, we calculate collective modes associated with fermion-pairs and bosons in a superfluid phase. We derive a stability condition of the mixtures, which is valid in the entire region of the BCS-BEC crossover. This stability condition, which smoothly connects the well-known results obtained in both the BCS and BEC limits, shows that sufficiently strong FB interactions destabilize the mixtures. In order to investigate the consequence of this instability, we study the ground state energy of both the uniformly mixed and the phase-separated states. We find that the FB repulsion induces a phase separation, whereas the FB attraction should cause a collapse of the mixture.
The possibilities for the experimental observation of such instabilities are also discussed.
\end{abstract}

\maketitle

\section{\label{sq:introduction}Introduction}

Recent experimental studies on ultracold fermions have provided new opportunities to study a crossover between a BCS superfluid and a molecular Bose-Einstein condensate (BEC). In such a system the strength of the two-body interaction can be freely tuned. 
In the weak two-body attraction limit, fermions form Cooper pairs and undergo the BCS transition.  
On the other hand, in the strong attraction limit, tightly bound pairs undergo Bose-Einstein condensation.  
Since these two limits are essentially understood as a condensation of the fermion pairs, a smooth crossover from a BCS state to a BEC state is expected to occur (BCS-BEC crossover).

The study of BCS-BEC crossover was pioneered by Eagles~\cite{eagles} and Leggett~\cite{leg}
and later extended to finite temperatures by Nozi\`eres and Schmitt-Rink~\cite{noz}. 
It was further extended by S\`a de Melo, Randeria, and Engelbrecht using functional integral formalism~\cite{melo-gl}.
Recent progress on the experimental techniques with the use of Feshbach resonances allows a realization of the BCS-BEC crossover.
The crossover from BCS to tightly bound molecular BEC was experimentally confirmed by the observation of pairing gap~\cite{chin-gap}, condensation fraction~\cite{zwi-cond,regal-cond}, heat capacity~\cite{kinast-heat}, collective oscillations~\cite{barten-co,kinast-co}, vortex lattices~\cite{zwi-vortex} and density profiles with a population imbalance~\cite{zwi-sf}.

Ultracold fermion systems have been studied with Bose-Fermi mixtures in association with fermion cooling. 
Cooling of degenerate fermions was difficult since the inter-fermion collisions are suppressed due to the Pauli exclusion principle.
However, by adding bosons, one can perform the sympathetic cooling~\cite{timm-cool} which yields realization of ultracold fermions.
In these systems, it is known that uniform Bose-Fermi mixtures become unstable for sufficiently strong fermion-boson (FB) interactions~\cite{mo-phase,roth-stab,roth-stab2,viv-phase,yi-phase-bf}; phase separation (collapse) occurs with repulsive (attractive) FB interactions. Experimentally, the collapse of a Bose-Fermi mixture was observed in a $^{40}\mathrm{K}$-$^{87}\mathrm{Rb}$ mixture~\cite{modu-collapse}. 
Therefore, experimental studies on ultracold fermions have been conducted to avoid the phase separation or the collapse of the mixture in order to perform the sympathetic cooling~\cite{str-molecule,mit-deg}.
Nevertheless, these instabilities themselves should have interesting features, thus they deserve an independent study.

In this paper, we investigate a Bose-Fermi mixture with a Feshbach resonance between two hyperfine states of fermions at absolute zero temperature. This system was recently studied by Salasnich and Toigo from the BCS to the unitarity region~\cite{salasnich}. We present a theory which is applicable in the whole region of the BCS-BEC crossover. Using a functional integral method, we calculate the collective modes associated with the quantum fluctuations of the boson and fermion-pair condensates.
Assuming that the bosons and the fermions are uniformly mixed, we find that the Bogoliubov mode of the bosons and the Anderson-Bogoliubov mode of the fermions are mixed by the FB interactions. 
In order to investigate the system more precisely, we derive a set of coupled Gross-Pitaevskii equations for the boson and fermion-pair condensates in the BEC limit, and show that the system is equivalent to two-component Bose gases. 
We also derive a stability condition against small quantum fluctuations. In order to understand the physical meanings of this condition, the ground state energy in the whole region of the crossover is derived. Using that, it is shown that the instability relates to the phase separation for repulsive FB interactions whereas the collapse for attractive FB interactions.
Especially when the system is initially in the uniformly mixed state, we have shown that the instability against small fluctuations corresponds to the phase separation for FB repulsion.
We also discuss how these instabilities can be observed in experiments. We expect $^6$Li is the best candidate to see such instabilities.

This paper is organized as follows.
In Sec.~\ref{sq:model}, an effective action is derived by performing a Hubbard-Stratonovich transformation.
In Sec.~\ref{sq:pert}, an effective action is derived by a second order perturbation expansion with respect to the quantum fluctuations associated with the boson and fermion-pair order parameters.
In Sec.~\ref{sq:collective}, collective modes are studied. The Bogoliubov mode of the bosons and the Anderson-Bogoliubov mode of the fermions are shown to be mixed by the FB interactions. In Sec.~\ref{sq:st}, a stability condition of the mixture is derived and discussed. In Sec.~\ref{sq:discussion}, the stability condition and the possibility of the experimental realization are discussed.

\section{\label{sq:model}Model}
We study the system at sufficiently low temperature, where both bosons and fermions are in a superfluid phase. We start with an action for a three-dimensional uniform Bose-Fermi mixture
\begin{equation}
	S=S_B+S_F+S_I,
\end{equation}
where $S_B$ and $S_F$ are the actions for  bosons and fermions,
\begin{eqnarray}
	S_B[\phi_B]
	&=&
	\int_0^{\beta} d\tau \int d \bm{x}
	\nonumber\\
	&&
	\bigg[\phi_B^\ast(\bm{x},\tau) \bigg(\frac{\partial}{\partial \tau} 
	-\frac{ \nabla^2}{2 m_B} - \mu_B \bigg) \phi_B(\bm{x},\tau)
	\nonumber\\
	&&
	+\frac{g_B}{2}| \phi_B(\bm{x},\tau) |^4 \bigg],
\end{eqnarray}
\begin{eqnarray}
	S_F[\psi_{\uparrow},\psi_{\downarrow}]&=& \sum_{\sigma=\uparrow\downarrow} 
	\int_0^{\beta} d\tau \int d \bm{x}
	\nonumber\\
	&&
	\bigg[ \psi^\ast_\sigma(\bm{x},\tau) \bigg(\frac{\partial}{\partial \tau} 
	-\frac{ \nabla^2}{2 m_F} - \mu_F \bigg) \psi_\sigma(\bm{x},\tau) 
	\nonumber\\
	&&
	+\frac{\tilde{g}_F}{2} |\psi_\sigma(\bm{x},\tau) |^2  | \psi_{- \sigma } (\bm{x},\tau) |^2 
	\bigg],
\end{eqnarray}
while $S_I$ represents the two-body FB interaction,
\begin{eqnarray}
	S_I[\phi_B,\psi_{\uparrow},\psi_{\downarrow}] 
	&=& \sum_{\sigma=\uparrow\downarrow} \int_0^{\beta} d\tau \int d \bm{x} 
	\nonumber\\
	&&
	g_I |\psi_\sigma(\bm{x},\tau) |^2  |\phi_B(\bm{x},\tau) |^2
\end{eqnarray}
with $\beta$ being the inverse of the temperature. Here, $\phi_B(\bm{x},\tau)$ is a complex field for bosons whereas $\psi_\sigma(\bm{x},\tau)$ are Grassmann fields for fermions with spin $\sigma=\uparrow\downarrow$. We denote three-dimensional vectors with bold letters. $\mu_B$ and $\mu_F$ are the chemical potentials for bosons and fermions, respectively. For simplicity, we assume that the numbers of the spin-up and -down particles are equal.
We consider the boson-boson (BB) and the FB s-wave scatterings with coupling constants $g_B=4\pi a_B/m_B$ and $g_I=2\pi a_I/m_I$, respectively. $a_B$ and $a_I$ are the scattering lengths and $m_I=m_F m_B/(m_F+m_B)$ is the reduced mass.
Throughout the paper, we set $\hbar=k_B=1$.

The coupling constant for the fermion-fermion (FF) interaction can be controlled by a Feshbach resonance. The bare s-wave attraction, $\tilde{g}_F$ is ``renormalized'' as
\begin{eqnarray}
	\frac{1 }{\tilde{g}_F}=\frac{1}{ g_F}-\int \frac{d\bm{p}}{(2\pi)^3}\frac{m_F }{ \bm{p}^2},
\end{eqnarray}
where $g_F=4\pi a_F/m_F$ is the effective FF interaction with $a_F$ being the scattering length which can be tuned from $-\infty$ to $+\infty$. 

The interaction part of $S_F$ can be transformed using the usual Hubbard-Stratonovich transformation by introducing new complex fields $\Delta(\bm{x},\tau)$ and $\Delta^\ast(\bm{x},\tau)$. Integrating over the Grassmann fields, we obtain an effective action,
\begin{eqnarray}
	S_{\mathrm{eff}}[\Delta,\phi_B] &=& S_B[\phi_B]
	\nonumber\\
	&&+\int_0^{\beta} d\tau \int d \bm{x} 
	\bigg[-\frac{|\Delta(\bm{x},\tau)|^2 }{ \tilde{g}_F} 
	\nonumber \\
	&& 
	-\frac{1}{\beta V}\mathrm{tr}[\mathrm{ln}(\beta\mathcal{G}^{-1})]\bigg]
	\label{s_eff}
\end{eqnarray}
where $V$ is the volume of the system and $\mathcal{G}^{-1}$ is the inverse Nambu propagator,
\begin{eqnarray}
	\mathcal{G}^{-1} &=&
	-\frac{\partial}{\partial\tau}-\Big(-\frac{\nabla^2 }{ 2m_F}-\mu_F+g_I |\phi_B|^2\Big)\sigma_z
	\nonumber\\
	&&
	-\Delta(\bm{x},\tau)\sigma^+-\Delta^*(\bm{x},\tau)\sigma^-,
	\label{nambu}
\end{eqnarray}
where $\sigma^{\pm}=(\sigma_x \pm i \sigma_y)/2$ with $\sigma_{i}$ being the Pauli matrices.

\section{\label{sq:pert}Perturbation Expansion}

Assuming that the temperature is sufficiently low, two order parameters, $\phi_B(\bm{x},\tau)$ and $\Delta(\bm{x},\tau)$, can be separated into their averages and fluctuations.
\begin{eqnarray}
	\phi_B(\bm{x},\tau)=\phi_0+\phi(\bm{x},\tau), \\
	\Delta(\bm{x},\tau)=\Delta_0+\delta(\bm{x},\tau). 
	\label{eq:fluctuation}
\end{eqnarray}
Then, the Nambu propagator in Eq.~(\ref{s_eff}) can be expanded as follows,
\begin{eqnarray}
	&&\mathrm{tr}\big[\mathrm{ln}(\beta\mathcal{G}^{-1})\big]
	\nonumber \\
	=&&\mathrm{tr}\big[\mathrm{ln}\big(\beta \mathcal{G}_0^{-1}\big)\big]-\sum_{n=1}^{\infty} \mathrm{tr}\big[\frac{(\mathcal{G}_0 \Sigma)^n }{ n}\big],
	\label{eq:expansion-gsigma}
\end{eqnarray}
where the unperturbed Green functions $\mathcal{G}_0$ and the self energy $\Sigma$ are 
\begin{eqnarray}
\mathcal{G}_0^{-1} &=&
-\frac{\partial}{\partial\tau}-\Big(-\frac{\nabla^2 }{ 2m_F}-\mu_F+g_I |\phi_0|^2\Big)\sigma_z
	\nonumber\\
	&&
	-\Delta_0\sigma^+-\Delta_0\sigma^-, \\
	\Sigma &=& 
	g_I\left( \phi_0\left(\phi\left(\bm{x},\tau\right)+\phi^\ast\left(\bm{x},\tau\right)\right)+ |\phi\left(\bm{x},\tau\right)|^2\right)\sigma_z 
	\nonumber\\
	&& +\delta(\bm{x},\tau)\sigma^++\delta^\ast(\bm{x},\tau)\sigma^-.
\end{eqnarray}
Note that we have performed a unitary transformation on $\mathcal{G}_0$ and $\Sigma$ to set $\Delta_0$ real. From now on we use the momentum representation of $\mathcal{G}_{0}$, which are determined as 
\begin{eqnarray}
	\mathcal{G}_0(p) &=&
		\begin{pmatrix}
			\mathcal{G}_{0\downarrow}(p) & \mathcal{F}_{0}(p) \\
			\mathcal{F}_{0}(p) & \mathcal{G}_{0\uparrow}(p)
		\end{pmatrix}
		\nonumber\\
		&=&
		\frac{-1 }{ \omega^2+\xi_{\bm{p}}^2+\Delta_0^2} 
		\begin{pmatrix}
			i \omega +\xi_{\bm{p}} & \Delta_0 \\
			\Delta_0 & i \omega -\xi_{\bm{p}}
		\end{pmatrix},
	\label{green}
\end{eqnarray}
where $\xi_{\bm{p}}=\bm{p}^2/2m_F-\mu_F$ and $p=(\bm{p},i \omega)$.
We now expand the action to the second order in the fluctuations around the saddle point. The saddle point conditions give the gap equation and the Hugenholtz-Pines relation,
\begin{eqnarray}
	\frac{\Delta }{ \tilde{g}_F} &=& \frac{1 }{ \beta} \sum_{\omega} \int \frac{d\bm{p} }{ (2 \pi)^3} \mathcal{F}_{0}(p), 
	\label{eq:gap}
	\\
	\mu_B &=& g_B n_B +g_I n_{F\downarrow} +g_I n_{F\uparrow},
	\label{eq:hp}
\end{eqnarray}
where $n_B$ and $n_{F\sigma}$ are boson and fermion densities:
\begin{eqnarray}
	n_{F\downarrow}=n_{F\uparrow} &=& \frac{n_F}{2}=
	\frac{1}{\beta}\sum_{\omega}\int\frac{d\bm{ p }}{ (2 \pi)^3} \mathcal{G}_{0\downarrow}(p),
	\label{particle-number}
\end{eqnarray}
with $n_F=n_{F\uparrow}+n_{F\downarrow}$.
By solving the saddle point conditions, Eqs.~(\ref{eq:gap}) and~(\ref{eq:hp}), with the particle number equation, Eq.~(\ref{particle-number}), self-consistently, one can calculate the physical quantities of this system.

We now separate the Bose field and the pairing field to their real and imaginary parts
\begin{eqnarray}
	\phi(p) &=& \frac{1 }{ \sqrt{2}}\big(\phi_A(p)+i\phi_P(p)\big), \\
	\delta(p) &=& \frac{1 }{ \sqrt{2}}\big(\delta_a(p)+i\delta_p(p) \big),
\end{eqnarray}
where $\phi_A$, $\phi_P$, $\delta_a$ and $\delta_p$ are set to be real. 
Before proceeding with the expansion, we define RPA polarization bubbles, which emerge in the second order in the fluctuations. We define
\begin{eqnarray}
	h^+(p) &=&\frac{1 }{ \beta}\sum_{\omega}\int \frac{d\bm{q}}{(2\pi)^3} {\mathcal{G}_0}_{\downarrow}(q){\mathcal{G}_0}_{\uparrow}(q+p), 
	\nonumber\\
	h^-(p) &=&\frac{1 }{ \beta}\sum_{\omega}\int \frac{d\bm{q}}{(2\pi)^3} {\mathcal{G}_0}_{\downarrow}(q+p){\mathcal{G}_0}_{\uparrow}(q), 
	\nonumber\\
	g_{\sigma}(p) &=&\frac{1 }{ \beta}\sum_{\omega}\int \frac{d\bm{q}}{(2\pi)^3} {\mathcal{G}_0}_{\sigma}(q){\mathcal{G}_0}_{\sigma}(q+p), 
	\nonumber\\
	f(p) &=&\frac{1 }{ \beta}\sum_{\omega}\int \frac{d\bm{q}}{(2\pi)^3} {\mathcal{F}_0}(q){\mathcal{F}_0}(q+p), 
	\nonumber\\
	k_\sigma^+(p) &=&\frac{1 }{ \beta}\sum_{\omega}\int \frac{d\bm{q}}{(2\pi)^3} {\mathcal{G}_0}_{\sigma}(p){\mathcal{F}_0}(q+p), 
	\nonumber\\
	k_\sigma^-(p) &=&\frac{1 }{ \beta}\sum_{\omega}\int \frac{d\bm{q}}{(2\pi)^3} {\mathcal{G}_0}_{\sigma}(q+p){\mathcal{F}_0}(p). 
	\label{eq:bubble}
\end{eqnarray}
Using the notations defined above, the second order effective action is given by
\begin{eqnarray}
	S_{\mathrm{eff}}^{(2)} &=& \frac{1}{\beta}\sum_{\omega}\int \frac{d\bm{ p }}{ (2 \pi)^3}
	\Big[\mathrm{\Psi}_p^{\dag}
	\mathcal{K}^{-1}\mathrm{\Psi}_{-p}\Big],
	\label{eq:s_eff2nd}
\end{eqnarray}
where $\mathrm{\Psi}_p=1/\sqrt{2}
	\begin{pmatrix}
		\phi_A(p) & \phi_P(p) & \delta_a(p) & \delta_p(p)  
	\end{pmatrix}^T$
is a vector notation of the fluctuation fields
and $\mathcal{K}^{-1}$ is a 4$\times$4 matrix
\begin{eqnarray}
	&&\mathcal{K}^{-1} =
	\begin{pmatrix}
		\frac{\bm{p}^2 }{ 2m_B} +2g_B n_B+K_{AA} &
		 \omega &
		K_{Aa} &
		i K_{Ap} 
		\\
		- \omega&
		\frac{\bm{p}^2 }{ 2m_B}&
		0 &
		0 
		\\
		K_{Aa}&
		0&
		K_{aa}&
		i K_{ap}
		\\
		-i K_{Ap}&
		0&
		-i K_{ap}&
		K_{pp}
	\end{pmatrix} \nonumber\\
	\label{2ndmat}
\end{eqnarray}
with
\begin{eqnarray}
	K_{AA} &=& g_I^2 n_B\big(2g_\downarrow(p)+2g_\uparrow(p)-4f(p)\big),\nonumber\\
	K_{Aa} &=& g_I\sqrt{n_B}\big(k_\downarrow^+(p) +k_\downarrow^-(p) -k_\uparrow^+(p)-k_\uparrow^-(p)\big),\nonumber\\
	K_{Ap} &=& g_I\sqrt{n_B}\big(k_\downarrow^+(p)-k_\downarrow^-(p) +k_\uparrow^+(p)-k_\uparrow^-(p)\big),\nonumber\\
	K_{aa} &=& -\frac{1}{ \tilde{g}_F}+\frac{h_\downarrow(p)+h_\uparrow(p) }{ 2}+f(p),\nonumber\\
	K_{pp} &=& -\frac{1}{ \tilde{g}_F}+\frac{h_\downarrow(p)+h_\uparrow(p) }{ 2}-f(p),\nonumber\\
	K_{ap} &=& \frac{h_\downarrow(p)-h_\uparrow(p) }{ 2}.
\end{eqnarray}

\section{\label{sq:collective}Collective Modes: Anderson-Bogoliubov Mode and Bogoliubov Mode}

\begin{figure}[htb]
\includegraphics{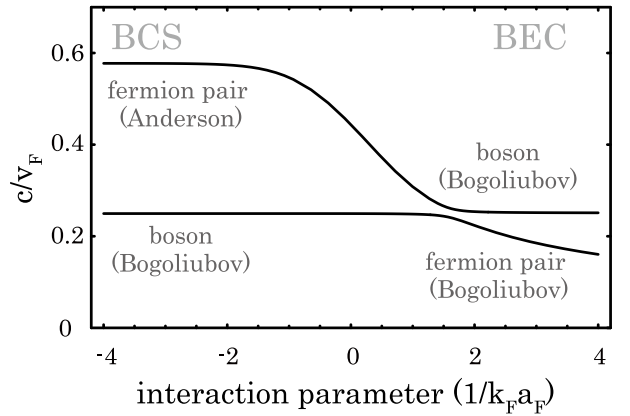}
\caption{\label{fig:sound} 
Sound velocities of the Bose-Fermi mixture in unit of $v_F$, the Fermi velocity in the BCS limit, calculated with  $m_B/m_F=7/6$, $ k_F a_I=0.2$, $ k_F a_B=0.4$ and $n_B/n_{F\sigma}=0.1$ with $k_F$ being the Fermi wave number in the BCS limit. 
}
\end{figure}

We now calculate the dispersion relation of the mixture using Eq.~(\ref{eq:s_eff2nd}). Assuming that the frequency is linear to the momentum in the long-wavelength limit $\omega=c p$, the RPA bubbles are expanded by $\omega$ and $p$ to the second order. After performing analytical continuation $i \omega\rightarrow\omega+i \eta $, the sound velocities $c$ are determined by solving $\mathrm{det}[\mathcal{K}^{-1}]=0$. The results are plotted in Fig.~\ref{fig:sound}. One can see that the collective excitations of the bosons and fermions exhibit coupled oscillations.

In both the BEC and the BCS limits, the dispersion relation is determined analytically. In the BCS limit, the dispersion relation reads
\begin{eqnarray}
	\omega^2 &&= \frac{1}{ 2} \bigg[ \Big( 
	\frac{g_B n_B }{ m_B} + \frac{v_F^2 }{ 3} \Big)
	\nonumber\\
	&&\pm
	\sqrt{  \Big(\frac{g_B n_B }{ m_B} - \frac{v_F^2 }{ 3} \Big)^2+\frac{8g_I^2 N(0)v_F^2n_B }{ 3 m_B} }
	\bigg]\bm{p}^2,
	\label{eq:velocity_BCS}
\end{eqnarray}
where $v_F=k_F/m_F$ is the Fermi velocity and $N(0)=m_F k_F/2\pi^2$ is the density of states of the fermions at the Fermi surface with $k_F = (3 \pi^2 n_F)^{1/3}$.
Two branches in Eq.~(\ref{eq:velocity_BCS}) corresponds to the Bogoliubov mode ($c_B=\sqrt{g_B n_B /m_B}$) of the bosons and Anderson mode ($c_A=v_F/\sqrt{3}$) of the fermions in the limit $g_I \to 0$; for $c_B > c_A$ ($c_B < c_A$), the positive and negative (negative and positive) signs correspond the Bogoliubov mode and Anderson mode, respectively.

In the BEC limit, the dispersion relation is calculated as
\begin{eqnarray}
	\omega^2 &=& \frac{1}{2}\bigg[
	\epsilon_B(\epsilon_B+2g_B n_B)+\epsilon_C(\epsilon_C+2g_C n_C)
	\nonumber\\
	&& \pm 
	\Big(\big(\epsilon_B(\epsilon_B+2g_B n_B)-\epsilon_C(\epsilon_C+2g_C n_C)\big)^2
	\nonumber\\
	&&+64g_I^2\epsilon_B\epsilon_C n_B n_C\Big)^\frac{1}{2}
	\bigg],
	\label{disp-bec}
\end{eqnarray}
where $\epsilon_B=\bm{p}^2/2m_B$, $\epsilon_C=\bm{p}^2/4m_F$ and $n_C=\Delta_0^2m_F^2a_F/8\pi=n_F/2$ is a fermion-pair density which will be discussed later. 
In the long-wavelength limit, Eq.~(\ref{disp-bec}) reduces to
\begin{eqnarray}
	\omega^2 &&= \frac{1 }{ 2}\bigg[
	\frac{g_B n_B}{ m_B}+\frac{g_F n_C }{  m_C} 
	\nonumber\\
	&&
	\pm \sqrt{\bigg(\frac{g_B n_B}{ m_B}-\frac{g_F n_C }{  m_C}\bigg)^2+\frac{16g_I^2n_B n_C}{ m_B m_C}}
	\bigg]\bm{p}^2,
	\label{eq:velocity_BEC}
\end{eqnarray}
which, in the limit $g_I\to 0$, corresponds to the Bogoliubov modes of the bosons and the fermion-pairs  with the mass $m_C = 2m_F$, inter-pair scattering length $2 a_F$ and pair condensate density $n_C=n_F/2$.

The result in the BEC limit can be understood more naturally by deriving a set of coupled Gross-Pitaevskii equations of the boson and fermion-pair condensates. Here, we employ a similar approach to the one in Refs.~\cite{melo-gl,pieri-gp} in order to depict the tightly bound fermion pair.
In the BEC limit, $\mu_F$ is the largest scale of the energies since fermions are tightly bound. This enables us to expand Eqs.~(\ref{eq:gap}) and (\ref{particle-number}) in terms of $\Delta/|\mu_F| \propto (k_F a_F)^{3/2}$, which is a small parameter in the BEC limit. One finds $\mu_F$ and $\Delta$ to be
\begin{eqnarray}
	2\mu_F &\sim& -\frac{1}{m_F a_F^2}+\frac{g_F n_F}{2}+2 g_I n_B,
	\label{eq:mu-gp}
	\\
	|\Delta|^2 &\sim& \frac{4 \pi n_F}{m_F^2 a_F}.
	\label{eq:gap-gp}
\end{eqnarray}
We now employ the off-diagonal part of the Dyson equation of the fermion Green function by taking a variation ${\delta S_{\mathrm{eff}}}/{\delta \Delta^\ast} =0$. We choose $\mathcal{G}_0$ as a Green function of ideal fermions and the self energy part $\Sigma$ as 
\begin{eqnarray}
	\Sigma = g_I |\phi(\bm{x},\tau)|^2 \sigma_z +\Delta(\bm{x},\tau)\sigma^++\Delta^\ast(\bm{x},\tau)\sigma^-.
	\label{eq:sigma-gp}
\end{eqnarray}
Then, the expansion of the Dyson equation by the self energy to the third order gives
\begin{eqnarray}
	&&
	-\frac{\Delta(\bm{x},\tau)}{\tilde{g}_F} 
	+\mathcal{G}_{0\downarrow}(\bm{x},\tau)\mathcal{G}_{0\uparrow}(\bm{x},\tau)\Delta(\bm{x},\tau) \nonumber\\
	&&+
	\mathcal{G}_{0\downarrow}(\bm{x},\tau)^2\mathcal{G}_{0\uparrow}(\bm{x},\tau)
	g_I |\phi(\bm{x},\tau)|^2\Delta(\bm{x},\tau) \nonumber\\
	&&-
	\mathcal{G}_{0\downarrow}(\bm{x},\tau)\mathcal{G}_{0\uparrow}(\bm{x},\tau)^2 
	g_I |\phi(\bm{x},\tau)|^2\Delta(\bm{x},\tau) \nonumber\\
	&&+
	\mathcal{G}_{0\downarrow}(\bm{x},\tau)^2\mathcal{G}_{0\uparrow}(\bm{x},\tau)^2
	|\Delta(\bm{x},\tau)|^2\Delta(\bm{x},\tau) 
	\nonumber \\
	&&=0. 
	\label{eq:dyson-expansion}
\end{eqnarray}
Note that we have assumed that $g_{I}|\phi|^2\ll\Delta$ and have neglected the second order contributions of $g_{I}|\phi|^2$.
The first term in Eq.~(\ref{eq:dyson-expansion}) can be calculated as
\begin{eqnarray}
	&&-\frac{\Delta(\bm{x},\tau)}{\tilde{g}_F} 
	+\mathcal{G}_{0\downarrow}(\bm{x},\tau)\mathcal{G}_{0\uparrow}(\bm{x},\tau)
	\Delta(\bm{x},\tau) \nonumber\\
	&&=-\frac{\Delta(\bm{x},\tau)}{g_F}
	+\frac{1}{\beta}\sum_{\nu}\int \frac{d\bm{ q}}{(2\pi)^3}e^{-i q x} \nonumber\\
	&&
	\int \frac{d\bm{ p}}{(2\pi)^3}
	\left[
	\frac{m_F}{\bm{p}^2}+
	\frac{1}{\beta}\sum_{\omega} 
	\mathcal{G}_{0\downarrow}(p)\mathcal{G}_{0\uparrow}(q+p)
	\right]  \Delta(q)
	\nonumber\\
	&&\sim
	-\frac{\Delta(\bm{x},\tau)}{g_F}
	+\frac{1}{\beta}\sum_{\nu}\int \frac{d\bm{ q}}{(2\pi)^3}e^{-i q x} \nonumber\\
	&&~~\left[\frac{m_F^{3/2}}{8\sqrt{2}\pi|\mu_F|^{1/2}}\left(
	-i \nu +\frac{\bm{q}^2}{4m_F}-4\mu_F
	\right)  \right] \Delta(q). \nonumber\\
	\label{eq:kernel-2}
\end{eqnarray}
Similarly, the remaining terms can also determined as
\begin{eqnarray}
	&&
	\mathcal{G}_{0\downarrow}(\bm{x},\tau)^2\mathcal{G}_{0\uparrow}(\bm{x},\tau)
	g_I |\phi(\bm{x},\tau)|^2\Delta(\bm{x},\tau) \nonumber\\
	&&-
	\mathcal{G}_{0\downarrow}(\bm{x},\tau)\mathcal{G}_{0\uparrow}(\bm{x},\tau)^2 
	g_I |\phi(\bm{x},\tau)|^2\Delta(\bm{x},\tau) \nonumber\\
	&&\sim
	\frac{m^{3/2}}{4\sqrt{2}\pi|\mu_F|^{1/2}}g_I |\phi(\bm{x},\tau)|^2\Delta(\bm{x},\tau),
	\label{eq:kernel-3}\\
	&&
	\mathcal{G}_{0\downarrow}(\bm{x},\tau)^2\mathcal{G}_{0\uparrow}(\bm{x},\tau)^2 |\Delta(\bm{x},\tau)|^2\Delta(\bm{x},\tau)\nonumber\\
	&&\sim
	\frac{m^{3/2}}{32\sqrt{2}\pi|\mu_F|^{3/2}}
	|\Delta(\bm{x},\tau)|^2\Delta(\bm{x},\tau).
	\label{eq:kernel-4}
\end{eqnarray}
We now define a new order parameter which corresponds to the fermion-pair condensate density
\begin{equation}
	|\Psi_C(\bm{x},\tau)|^2 
	= \frac{m_F^2 a_F}{8\pi}|\Delta(\bm{x},\tau)|^2
	\sim \frac{n_F}{2}=n_C.
	\label{eq:new-order}
\end{equation}
Substituting Eqs.~(\ref{eq:kernel-2}-\ref{eq:new-order}) to Eq.~(\ref{eq:dyson-expansion}), and using Eqs.~(\ref{eq:mu-gp}) and (\ref{eq:gap-gp}), one obtains the Gross-Pitaevskii equation of the fermion-pair condensates
\begin{eqnarray}
	&&
	i \frac{\partial}{\partial t}\Psi_C
	-\frac{\nabla^2}{4m_F}\Psi_C
	+g_F|\Psi_C|^2\Psi_C
	\nonumber\\
	&&
	+2g_I|\phi|^2\Psi_C
	=\mu_C \Psi_C,
	\label{GP_C}
\end{eqnarray}
where $\mu_C=2\mu_F+1/m_F a_F^2$ is the effective chemical potential of the fermion-pairs. From $\delta S_{\mathrm{eff}}/\delta\phi=0$, we can also derive the Gross-Pitaevskii equation for the bosons,
\begin{eqnarray}
	&&i \frac{\partial}{\partial t}\phi
	-\frac{\nabla^2}{2m_B}\phi+g_B|\phi|^2\phi
	\nonumber\\
	&&
	+2g_I|\Psi_C|^2\phi
	=\mu_B \phi.
	\label{GP_B}
\end{eqnarray}
From Eqs.~(\ref{GP_C}) and~(\ref{GP_B}), one can derive a set of coupled Bogoliubov de Gennes equations, and reproduce the collective modes of Eq.~(\ref{disp-bec}). This fact implies that, in the BEC limit, the system is equivalent to the two-component Bose gases with the atomic mass of $m_B$ and $2m_F$ and the scattering length of $a_B$ and $2a_F$. Note that the coupling constant for boson fermion-pair interaction is $2g_I=4\pi a_I(m_F+m_B)/m_Fm_B$ rather than $4 \pi a_I (2m_F+m_B)/m_Fm_B$ since the FB interactions essentially act between one fermion and one boson.

\section{\label{sq:st}Stability Condition}

For sufficiently strong FB interactions, the sound velocity of the lower energy branch becomes a pure imaginary, which implies the instability of the system. Then, the stability boundary is determined by the condition that one of the sound velocities is zero. Then, the stability condition is expressed with the RPA bubbles, 
$k(0)=k_{\downarrow}^{\pm}(0)=-k_{\uparrow}^{\pm}(0)$, as
\begin{eqnarray}
	g_B / g_I^2 &>& 4f(0)+4{k(0)}^2/f(0).
	\label{eq:stability}
\end{eqnarray}

\begin{figure}[htb]
\includegraphics{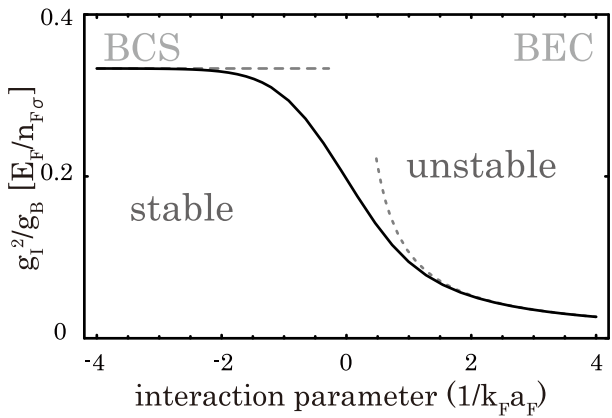}
\caption{\label{fig:stability} 
Stability condition for Bose-Fermi mixtures plotted from Eq.~(\ref{eq:stability}) (solid line). The dashed and dotted line corresponds to the BCS limit, Eq.~(\ref{eq:stab_BCS}) and the BEC limit, Eq.~(\ref{eq:stab_BEC}), respectively.
}
\end{figure}

In the BCS limit, Eq.~(\ref{eq:stability}) becomes 
\begin{eqnarray}
	g_B/g_I^2 > 2N(0),\label{eq:stab_BCS}
\end{eqnarray}
which agrees with the thermodynamic stability condition for Bose-Fermi mixtures~\cite{stoof,viv-ground,viv-phase,yip-collective,valtan,yi-phase-bf}. 
On the other hand, in the BEC limit, Eq.~(\ref{eq:stability}) reduces to 
\begin{eqnarray}
	g_F g_B > 4 g_I^2, \label{eq:stab_BEC}
\end{eqnarray}
which again coincides with the thermodynamic stability condition for two-component Bose gases~\cite{two-components-bec,esry-two-comp,comment-BEC}. 
These results are summarized in Fig.~\ref{fig:stability}. One can see Eq.~(\ref{eq:stability}) smoothly connects the well-known stability conditions of the Bose-Fermi mixtures Eq.~(\ref{eq:stab_BCS}) and two-component Bose gases Eq.~(\ref{eq:stab_BEC}).

Till now, we have assumed that the ground state of the mixture is a uniformly mixed state. In order to understand the consequence of the instability, we study the free energies of both the uniformly mixed states and the phase-separated states.
At the zero temperature, we can approximate the thermodynamic potential as $\Omega=-P V \sim \beta^{-1} S^{(0)}$ where $P$ is the pressure, $V$ is the volume and $S^{(0)}$ is the effective action without the quantum fluctuations. 
Using Eqs.~(\ref{s_eff}) and (\ref{eq:expansion-gsigma}), and then setting $\mathcal{G}_{0}$ as the Green function of the ideal fermions and $\Sigma=g_I |\phi_0|^2\sigma_z+\Delta_{0}\sigma^++\Delta_{0}^*\sigma^-$, the free energy of the uniformly mixed state is determined to be
\begin{eqnarray}
	\mathcal{E}_{\mathrm{mix}} &=&
	-P_F^{(0)}(N_F,V)V-P_F^{\mathrm{(sc)}}(N_F,V)V
	\nonumber\\
	&&
	+\mu_F(N_F,V) N_F 
	+\frac{g_B }{ 2}\frac{N_B^2 }{ V}
	+{g_I}\frac{N_B N_F }{ V}, 
\end{eqnarray}
where
\begin{eqnarray}
	P_F^{(0)} &=& 
	\frac{2 }{ 5}\frac{(2m_F \mu_F)^{\frac{3 }{ 2}}}{ 3\pi^2} \mu_F \theta(\mu_F),
	\label{fermiP-0}
	\\
	P_F^{\mathrm{(sc)}} &=& 
	\frac{ {|\Delta_0|}^2 }{ g_F} 
	+\int \frac{d\bm{ p }}{ (2 \pi)^3}
	\nonumber\\
	&& \bigg(\sqrt{\xi_{\bm{p}}+|\Delta_0|^2}-|\xi_{\bm{p}}|-\frac{|\Delta_0|^2 m_F}{  \bm{p}^2} \bigg)
	\label{fermiP-sc}
\end{eqnarray}
with $N_B=n_{B} V$  and $N_F=n_{F} V$. Note that we have neglected the higher order contributions of the FB interactions.

Similarly, one can also calculate the free energy of the phase-separated states as
\begin{eqnarray}
	\mathcal{E}_{\mathrm{sep}} &=&
	-P_F^{(0)}(N_F,V_F) V_F -P_F^{\mathrm{(sc)}}(N_F,V_F) V_F 
	\nonumber\\
	&&
	+\mu_F(N_F,V_F) N_F
	+ \frac{g_B }{ 2}\frac{N_B^2}{ V_B},
	\label{ene-sep}
\end{eqnarray}
with $V_F$ and $V_B$ being the volume occupied by fermions and bosons, respectively.

\begin{figure}[htb]
\includegraphics{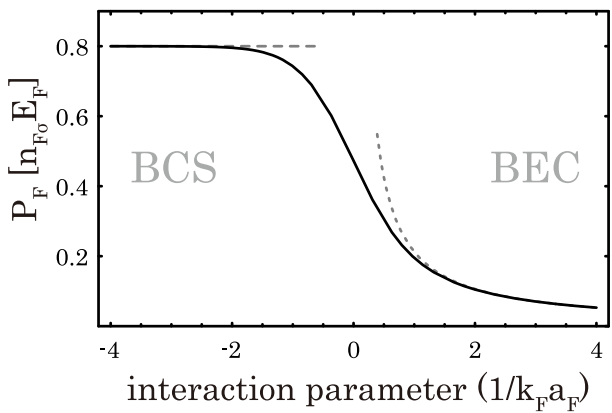}
\caption{\label{fig:pressure} 
Fermi pressure $P_{F}^{(0)}+P_{F}^{\mathrm{(sc)}}$ plotted against interaction parameter (solid line). In the BCS limit, it approaches to $2/5\mu_F n_F$ (gray dashed line) whereas in the BEC limit, it approaches $g_F n_{C}^2/2$ (gray dotted line).}
\end{figure}

We now show that the stability condition Eq.~(\ref{eq:stability}), relates to the condition for the system to be in a uniformly mixed state rather than a phase-separated state. In order for phase separation to occur, the following two conditions must be satisfied; the equilibrium condition for the pressure, 
\begin{eqnarray}
	P_F^{(0)}+P_F^{\mathrm{(sc)}}=\frac{g_B}{2} \frac{N_B^2}{ V_B^2}.
	\label{eq:pressure-eq}
\end{eqnarray}
and the condition for the free energies,
\begin{equation}
	\mathcal{E}_{\mathrm{sep}} < \mathcal{E}_{\mathrm{mix}}.
	\label{eq:phase-separation}
\end{equation}
In the BEC limit, $P_F^{(0)}$, the pressure of the ideal Fermi gas, vanishes because of the disappearance of the Fermi surface. Expanding $P_F^{\mathrm{(sc)}}$ by $\Delta_0/|\mu_F|$, which is a small parameter in the BEC limit, $P_F^{\mathrm{(sc)}}\to g_F n_C^2/2$ as shown in Fig.~\ref{fig:pressure}. Remembering that the chemical potential of the fermions becomes $2\mu_F\to-1/m_F a_F^2 + g_F n_C$, the condition for phase separation becomes
\begin{eqnarray}
	\mathcal{E}_{\mathrm{sep}}-\mathcal{E}_{\mathrm{mix}} &=& 
	\frac{g_F}{2}\frac{ N_C^2}{ V_F}+ \frac{g_B }{ 2}\frac{ N_B^2}{ V_B}
	\nonumber\\
	&&
	-\frac{g_B}{2}\frac{ N_B^2}{ V}-\frac{g_F}{2}\frac{ N_C^2}{ V}-2g_I\frac{ N_B N_C}{ V} 
	\nonumber\\
	&<&0,
	\label{eq:phase-separation-condition}
\end{eqnarray}
where $N_C=N_F/2$ is a fermion pair number.
Using the equilibrium condition Eq.~(\ref{eq:pressure-eq}), which reduces to $g_F N_C^2/2V_F^2=g_B N_B^2/V_B^2$, the condition for phase separation in the BEC limit becomes
\begin{eqnarray}
	4g_I^2 > g_B g_F
	\label{phase-separation-bec}
\end{eqnarray}
for repulsive FB interactions. Eq.~(\ref{phase-separation-bec}) agrees with the condition for phase separation of the two-component Bose gases~\cite{pita,esry-two-comp}. Phase separation of the two-component Bose gases was actually observed in the experiment~\cite{hall-two-bec-sep}. This condition also agrees with the stability condition against the small fluctuations in the BEC limit, Eq.~(\ref{eq:stab_BEC}). 

In the BCS limit, $P_F^{\mathrm{(sc)}}$ is negligible compared to $P_F^{\mathrm{(0)}}$ since $\Delta_0/\epsilon_F \ll 1$. Then the argument on the Bose-Fermi mixture with one-component fermions also applies to our case. Using the results of the previous work~\cite{viv-phase,comment-phase}, which basically compares the ground state energies of the uniformly mixed states and the phase-separated states, Eq.~(\ref{eq:stab_BCS}) corresponds to the condition that the system does not have any phase-separated states in the dilute boson density limit. 

Therefore, with repulsive FB interactions, Eq.~(\ref{eq:stability}) exactly corresponds to the condition for phase separation to occur in the BEC limit, whereas it corresponds to the condition only in the limiting case in the BCS limit.
However, in the next section, we will show that  the phase diagram is determined by Eq.~(\ref{eq:stability}) regardless of the density of bosons if the system is initially in a uniformly mixed state.

We point out that Salasnich and Toigo also found the stability conditions based on the approximation of the ground state energy calculated by a Monte Carlo study \cite{salasnich}. They explicitly calculated the stability conditions which agree with our findings in the BCS limit. However, their approach is currently limited in the BCS to the unitarity region. In contrast, our theory is also valid in the BEC region.

When the FB interactions are attractive, the condition Eq.~(\ref{eq:phase-separation}) is not satisfied since an increase of $|g_I|$ lowers the ground state energy of the uniformly mixed state.  Therefore, the phase-separated state is unstable. In this case, we expect the mixture to collapse by the FB attraction. 

\section{\label{sq:discussion}Discussion}

In this section, we discuss how our stability condition can be verified in real systems. We point out that the phase diagram is governed by the stability condition Eq.~(\ref{eq:stability}) assuming that the system is initially in the uniformly mixed state. 

The stability condition is derived so that the system is robust against the fluctuations. 
In other words, it is the condition that a uniformly mixed ground state is at the local minimum of the free energy.
Consider the case that the interaction parameter is swept adiabatically from the BCS to the BEC region. At sufficiently low temperatures, there is no process for the system to make a transition to the other state. 
Therefore, the system remains at its initial state, even when a non uniformly-mixed state, e.g. the phase-separated state, is energetically favored due to the confining potential.
However, once the condition Eq.~(\ref{eq:stability}) is violated, the system becomes unstable against any small fluctuations.
Therefore, its phase diagram is expected to be like Fig.~\ref{fig:stability} when the system is initially in a uniformly mixed state. 

This explains the contradiction between our results and that of the experiment of the $^{40}\mathrm{K}$-$^{87}\mathrm{Rb}$ mixture~\cite{modu-collapse}. In this experiment, the condition for the collapse depends on the densities of both fermions and bosons whereas Eq.~(\ref{eq:stability}) only depends on the fermion density. 
This can be explained by considering that the initial state of the experiment where the collapse was observed was not a uniformly mixed state. 
In this case, the stability condition should be governed by $\mathcal{E}_{\mathrm{other}}>\mathcal{E}_\mathrm{mix}$ where $\mathcal{E}_{\mathrm{other}}$ is the ground state energy of the state after the collapse.

We now discuss how one could experimentally demonstrate our results.
In order to demonstrate the stability condition Eq.~(\ref{eq:stability}), we first need to prepare a uniformly mixed state and then change some parameters to make the system unstable.
Presently, it seems difficult to control the FB interactions. However, one can still tune $a_F$ to make the system unstable. 
Hence, if a uniformly mixed state was successfully realized in the BCS or in the unitarity region, where the system is likely to be stable (see Fig.~\ref{fig:stability}), the transition to the other ground state can be observed by tuning $a_F$ to the BEC region.

Currently there are two types of systems in which the BCS-BEC crossover was observed~\cite{zwi-cond,regal-cond,kinast-heat,zwi-sf,zwi-vortex,barten-co,kinast-co,chin-gap}; $^{40}$K of the hyperfine states of $|f,m_f\rangle=|9/2,-7/2\rangle$ and $|9/2,-9/2\rangle$ and $^6$Li of $|1/2,1/2\rangle$ and $|1/2,-1/2\rangle$. As far as we know, there has been no FB interaction scattering length reported in these states. However, we can still estimate the magnetic field strength where the instability occurs using an approximate expression of $a_F$,
\begin{equation}
	a_F \sim a_{\mathrm{bg}}\bigg(1-\frac{\Delta B}{B-B_0}\bigg),
	\label{eq:scat}
\end{equation}
where $B$ is the magnetic field strength, $B_0$ is the magnetic field at the resonance, $a_{\mathrm{bg}}$ is the scattering length in the small or large $B$ limit and $|\Delta B|$ is corresponded to the width of the resonance. 

In the case of $^{40}$K, $a_{\mathrm{bg}}=174a_0$~\cite{regal-af} where $a_0$ is Bohr radius. Taking $a_{\mathrm{bg}}$ as the scattering length in the BEC limit, the range of the parameters where instability can be observed is $174a_0\le(1+\tilde{m}_B)^2a_I^2/(4\tilde{m}_Ba_B)$ with $\tilde{m}_B=m_B/m_F$. 

The observation of the transition is more promising in the case of $^6$Li since $a_{\mathrm{bg}}=-1405a_0$~\cite{bart-scat}. The system is in the pure BEC limit, i.e. $a_F=+0$ at $B=534\mathrm{G}$. Here, Eq.~(\ref{eq:stability}) automatically becomes false. Therefore, if one could realize a uniformly mixed state in the BCS or the unitarity region and tune the $a_F$ to the BEC region, one should see the instability for $B>543\mathrm{G}$. However, the experiment shows that the fermion-pair condensate vanishes around $B \sim 710\mathrm{G}$~\cite{zwi-cond}. Therefore, in order to see the transition such as phase separation and collapse, an appropriate hyperfine state which fails to meet Eq.~(\ref{eq:stability}) in $B>710\mathrm{G}$ must be chosen.

In experiments, mixtures are trapped by an external trap potential, which is not considered in our present homogeneous theory.
Assuming $N_B$ and $N_F \gg 1$, the effect of a trap can be reasonably investigated by performing a local density approximation, which modifies $\mu_F$ and $\mu_B$ to $\mu_F,\mu_B \rightarrow \mu_F,\mu_B-V_{\mathrm{ext}}(\bm{x})$ where $V_{\mathrm{ext}}(\bm{x})$ is the external trap potential. Normalization conditions, e.g. Eq.~(\ref{particle-number}), are also changed to 
$N_F=\beta^{-1}\int d \bm{x} \sum_{p}
\mathcal{G}_{0}\left(p;\mu_F-V_{\mathrm{ext}}(\bm{x})\right)$ 
and 
$N_B=\int d\bm{x}~n_B\left(\bm{x};\mu_B-V_{\mathrm{ext}}\left(\bm{x}\right)\right)$.
This approximation fails to describe the situation at the edge of the trap. This can be understood by considering the case that the same number of fermions and bosons are present in the same trap, and there is no FB interaction. Due to the Pauli exclusion principle, fermions are usually spread more than bosons, therefore, even though no FB interaction is present, phase separation can take place at the edge of the trap. 

However, this approximation well describes the physics at the center of the trap, where instability due to fluctuations is expected to occur first. The stability condition Eq.~(\ref{eq:stability}), which is found for a homogeneous case, is equivalent to the stability condition at the center of the trap which is described within a local density approximation. Therefore, we expect that our condition is still useful for trapped systems. However, it should be pointed out that for FB repulsion, the effect of instability may be less visible for the trapped case since the instability occurs only at the center of the trap. On the contrary, instability occurs simultaneously in the all region for the homogeneous case. For the FB attraction, instability at the center of a trap induces instability in the all region, therefore, experimentally, the phase diagram Fig.~\ref{fig:stability} could be observed more clearly.

\section{\label{sq:conclusion}Conclusion}

We have studied a system of a Bose-Fermi mixture with a Feshbach resonance between two hyperfine states of fermions in a superfluid phase. It has been shown that the Anderson-Bogoliubov mode and the Bogoliubov mode are mixed by FB interactions. We have also shown that the system becomes unstable as the strength of the FB interactions increases. 
The stability condition derived here has been shown to smoothly connect the well-known results of both the BCS and BEC limits.
In order to understand the stability condition, we have studied the ground state energy of this system for both the uniformly mixed and the phase-separated states. 
It has been shown that the stability condition relates to the condition for phase separation if the FB interactions are repulsive. 
Especially when the system is initially in the uniformly mixed state, we have shown that it corresponds to the condition for phase separation.
We have also shown that if a uniformly mixed Bose-Fermi mixture was prepared experimentally, the stability condition Eq.~(\ref{eq:stability}) can be verified. We expect $^6$Li is the best candidate for the transition to phase separation or collapse to be observed.

\begin{acknowledgments}
	We acknowledge Ippei Danshita for a fruitful discussion. We also acknowledge K. Iigaya, K. Kamide and D. Yamamoto for useful comments and supports. This work is supported by the 21st century COE Program (Holistic Research and Education Center for Physics of Self-organization Systems) at Waseda University from the Ministry of Education, Culture, Sports, Science and Technology of Japan.
\end{acknowledgments}
\newpage 


\begin{thebibliography}{35}
\bibitem[{com({\natexlab{a}})}]{comment-BEC}
\bibinfo{note}{In the BEC limit, the free energy is well approximated by
  $\mathcal{F}=g_Bn_B^2/2+g_Fn_C^2/2+2g_In_Bn_C$. The thermodynamic stablity
  condition, $\mathrm{det}[\partial^2 \mathcal{F}/\partial n_i \partial n_j]\ge
  0$ gives the same stablity condition, $g_Fg_B \ge 4g_I^2$}.
\bibitem[{com({\natexlab{b}})}]{comment-phase}
\bibinfo{note}{Since the work by Viverit~\cite{viv-phase} is a mixture of
  bosons and one-component fermions, the stability condition to the
  fluctuations, i.e. Eq. (\ref{eq:stab_BCS}), differs. We are working with two
  hyperfine states of fermions. Therefore, we have twice more density of
  states.}
\end{thebibliography}
\end{document}